\documentclass[twocolumn]{aastex63}
\usepackage{graphics,epsf}
\usepackage{amsmath}                
\usepackage{amsfonts}               
\usepackage{amssymb}                
\usepackage{epsfig}                 
\usepackage{appendix}
\usepackage{graphicx}
\usepackage{float}
\usepackage{color}
\usepackage{multirow}
\usepackage{colortbl}
\usepackage[para,online,flushleft]{threeparttable}

\hypersetup{citecolor=blue, 
            linkcolor=red, 
            menucolor=blue, 
            urlcolor=blue}  

\newcommand{\cm}{{~\rm cm}}
\newcommand{\m}{{~\rm m}}
\newcommand{\km}{{~\rm km}}
\newcommand{\s}{{~\rm s}}
\newcommand{\h}{{~\rm h}}

\newcommand{\g}{{~\rm g}}
\newcommand{\G}{{~\rm G}}
\newcommand{\K}{{~\rm K}}
\newcommand{\erg}{{~\rm erg}}
\newcommand{\yr}{{~\rm yr}}

\newcommand{\AU}{{~\rm AU}}

\newcommand{\days}{{~\rm days}}

\begin{document}

\title{Terminating a common envelope jets supernova impostor event with a super-Eddington blue supergiant}


\author[0009-0001-0720-6816]{Tamar Cohen}
\affiliation{Department of Physics, Technion, Haifa, 3200003, Israel; tamarco@campus.technion.ac.il; soker@physics.technion.ac.il}

\author[0000-0003-0375-8987]{Noam Soker}
\affiliation{Department of Physics, Technion, Haifa, 3200003, Israel; tamarco@campus.technion.ac.il; soker@physics.technion.ac.il}

\begin{abstract}
We conducted one-dimensional stellar evolutionary numerical simulations to build blue supergiant stellar models with a very low-envelope mass and a super-Eddington luminosity of $10^7 L_\odot$ that mimic the last phase of a common envelope evolution (CEE) where a neutron star (NS) accretes mass from the envelope and launches jets that power the system. Common envelope jets supernovae (CEJSNe) are CEE transient events where a NS spirals-in inside the envelope and then the core of a red supergiant (RSG) star accretes mass and launches jets that power the transient event. In case that the NS (or black hole) does not enter the core of the RSG the event is a CEJSN-impostor. We propose that in some cases a CEJSN-impostor event might end with such a phase of a blue supergiant lasting for several years to few tens of years.  The radius of the blue supergiant is about tens to few hundreds’ solar radii. We use a simple prescription to deposit the jets’ energy into the envelope. We find that the expected accretion rate of envelope mass onto the NS at the end of the CEE allows the power of the jets to be as we assume, $10^7 L_\odot$. Such a low-mass envelope might be the end of the RSG envelope, or might be a rebuilt envelope from mass fallback. Our study of a blue supergiant at the termination of a CEJSN-impostor event adds to the rich variety of transients that CEJSNe and CEJSN-impostors might form. 
\end{abstract}

\section{Introduction} 
\label{sec:intro}

In a common envelope jet supernova (CEJSN) event a neutron star (NS) or a black hole (BH) spiral-in inside the envelope of a red supergiant (RSG) star and then inside its core, accrete mass and launche jets (e.g., \citealt{Papishetal2015, SokerGilkis2018, Gilkisetal2019, GrichenerSoker2019a, LopezCamaraetal2019, LopezCamaraetal2020MN, Soker2021, Grichener2023}). The jets explode the star and power the transient event.  In this study we deal with the \textit{CEJSN impostor scenario} \citep{Gilkisetal2019}, where the NS/BH spirals-in inside the envelope of the RSG but does not merge with, and hence does not destroy, the core \citep{Gilkisetal2019}. 

The two opposite jets that the NS/BH launches as it accretes mass from the envelope collide with the outer envelope and with circumstellar material that early jets and/or the common envelope evolution (CEE) process have removed. The post-shock jet's gas and the post-shock ambient gas form a `cocoon' (e.g., \citealt{LopezCamaraetal2019}). Photons diffusing out from the cocoon form a bright transient event, a CEJSN (or CEJSN-impostor) event (e.g., \citealt{Schreieretal2021}) that mimics a peculiar core collapse supernova (CCSN; e.g., \citealt{SokeretalGG2019}). Examples include the  unusual gamma-ray burst GRB~101225A that \cite{Thoneetal2011} explain by a NS that merged with a helium star and the enigmatic supernovae iPTF14hls \citep{Arcavietal2017} and SN~2020faa \citep{Yangetal2020} that might be be CEJSN events (e.g., \citealt{SokerGilkis2018}). 
Adopting the CEJSN scenario and processes from these earlier studies, 
\cite{Dongetal2021} followed these studies and suggested the CEJSN scenario for the luminous radio transient VT~J121001+495647. CEJSNe might be the sites of some extreme-condition processes, like a fraction of the r-process nucleosynthesis in the Universe \citep{GrichenerSoker2019a, GrichenerSoker2019b, Gricheneretal2022, GrichenerSoker2022} and the site of some  fraction of the very high-energy neutrinos in cases of a BH companion that spirals-in inside the RSG envelope \citep{GrichenerSoker2021}. {{{{ Even before entering the CEE the NS/BH might accrete mass at a high rate and launch jets that produce very high-energy neutrino (\citealt{Sridharetal2022}; they term this phase hypernebula, \citealt{SridharMetzger2022}).    }}}}

The large energy of the jets that the NS/BH launches in CEJSNe is the main difference from many other binary interaction routes with and without CEE (e.g., \citealt{Hanetal2020}) and from the much less powerful jets that a main sequence companion might launch in a CEE (e.g., \citealt{Shiberetal2016}). 
The high energy-jets also distinguish the CEJSN scenario from related NS/BH scenarios of CEE without jets (e.g., \citealt{ThorneZytkow1977, FryerWoosley1998, ZhangFryer2001, BarkovKomissarov2011, Thoneetal2011, Chevalier2012, Schroderetal2020}). 

The peculiar transient AT2018cow \citep{Prenticeeta2018} and the CEJSN-impostor scenario for AT2018cow-like fast blue optical transients (FBOTs) motivate our study. AT2018cow-like FBOTS are bright transients, which might be brighter than superluminous CCSNe (e.g., \citealt{Marguttietal2019}), and with a rise time of only few days (e.g., \citealt{Prenticeeta2018, Perleyetal2019}). Their kinetic energy is $\simeq 10^{51}-10^{52} \erg$, they possess velocity components of up to $\ga 0.1 c$ (e.g., \citealt{Coppejansetal2020}), they might show hydrogen spectral lines (e.g., \citealt{Marguttietal2019}), they might display rapid X-ray variability (e.g., \citealt{Pashametal2022, Yaoetal2022}), have a dense circumstellar material (CSM; e.g., \citealt{NayanaChandra2021, Brightetal2021}), and tend to occur in star-forming galaxies (e.g., \citealt{Prenticeeta2018}).

There are several scenarios for AT2018cow-like FBOTs (e.g., \citealt{Liuetal2018, FoxSmith2019, Kuinetal2019, LyutikovToonen2019, Marguttietal2019, Quataertetal2019, Yuetal2019, Leungetal2020, Mohanetal2020, PiroLu2020, UnoMaeda2020, Kremeretal2021, Xiangetal2021, ChenShen2022, Gottliebetal2022}). \cite{SokeretalGG2019} proposed the polar CEJSN channel for AT2018cow. In this scenario early jets remove most of the envelope material along the polar directions, allowing late jets to expand almost freely to large distances, accounting for the fast rise and decline of FBOTs and for the $> 0.1c$ velocities. 

\cite{Soker2022FBOT} considers the polar CEJSN-impostor scenario for AT2018cow, i.e., a scenario where the NS companion does not destroy nor enter the core of the RSG. According to the CEJSN-impostor scenario many FBOTs become the progenitors of binary compact objects mergers (NS/BH-NS/BH merger). In this study we adopt this scenario that has four main phases (for schematic illustrations see figures in \citealt{Soker2022FBOT}). (1) A RSG star expands and tidally interacts with a NS/BH companion. The interaction enhances the mass loss rate of the RSG star to form a CSM. (2) Just before the onset of a CEE, months to few years before explosion, the NS/BH launches jets as it accretes mass. The jets inflate two opposite lobes at $\approx 100 \AU$ inside the CSM. (3) The binary system enters a CEE that ejects most of the envelope in a dense equatorial outflow. (4) The CEE ends with a circumbinary disk around the binary system of the NS/BH and the core of the RSG star. The NS/BH accretes mass from the circumbinary disk and launches energetic jets. The collision of the jets with the CSM in the two lobes powers an FBOT event. 
In a later time (long after the FBOT event) the core remnant of the RSG collapses to form a NS/BH in a CCSN explosion. 

In this study we consider another possible phase of evolution after the FBOT event, where the NS and the core orbit each other inside a {{{{ a newly rebuilt }}}} very low-mass common envelope.
{{{{ Namely, after the RSG envelope contracted to leave the NS outside the envelope at the termination of the main CEE phase, further evolution of the core and/or accretion of fallback material, {{{{{ most likely from the circumbinary disk before it disappears or from equatorial fallback beyond the circumbinary disk,  }}}}} caused the RSG envelope to re-expand and engulf the NS again. This leads to a second CEE phase, now with a very low mass envelope. {{{{{ We note that a secular evolution (stellar evolution that does not include extra ingredients) does not lead by itself to a second CEE phase with a NS, but rather to a stable mass transfer, Case BB mass transfer (e.g., \citealt{Taurisetal2015, Taurisetal2017, VignaGomezetal2018}).
To form a second CEE phase requires additional process(es), that might be rare. One such effect might be the same effect that causes some massive stars to experience outbursts years to months before they explode, i.e., strong core activity. Another possibility is that the rapid mass removal by the NS in the first CEE phase, due to the jets, causes the envelope to over-contracts. The envelope then expands in a time scale of months which is the thermal time of the outer envelope zones alone. \cite{SokerBear2023} find that the rapid removal of mass from a $7M_\odot$ star might reduce its radius to $2.1 R_\odot$ and that within about year it re-expands to $2.5 R_\odot$ with an expansion rate that rapidly decreases. We leave it to future studies to find the most likely scenario to form a second CEE.  }}}}}

The envelope density is too low to force the NS and the core to merge {{{{{ and practically the additional spiralling-in is negligible. }}}}}. However, the accretion energy onto the NS, via jets, powers this common envelope system for several years to tens of years. The motivation to propose this phase of evolution comes from the observation by  \cite{Sunetal2022} who detect a luminous source at $\simeq 2-3$ years after the outburst of AT2018cow that is consistent with a hot, $T> 5 \times 10^4 \K$, and luminous, $L > 10^7 L_\odot$, black body radiation source. {{{{{ In a very recent study \cite{Chenetal2023a} also study and discuss this very hot black-body source. }}}}} (for other recent studies of AT2018cow see, e.g., \citealt{Zhangetal2022, Chenetal2023b, Sunetal2023}). We describe this scenario in section \ref{sec:TheScenario}. In section \ref{sec:Numerical} we simulate a spherically symmetric stellar model with energy deposition that mimics the powering jets, and in section \ref{sec:Properties} we calculate the power of the jets to show self-consistency. We summarize this study is section \ref{sec:Summary}. 

\section{The scenario} 
\label{sec:TheScenario}

\cite{Sunetal2022} reported the detection of a luminous source  $\simeq 2-3$ years after the outburst of AT2018cow that is consistent with a hot, $T> 5 \times 10^4 \K$, and luminous, $L > 10^7 L_\odot$, black body radiation source. {{{{{ \cite{Chenetal2023a}  derive black-body temperatures and radii at $t\simeq 703 \days$ and $t\simeq 1453 \days$ of $(R,T) \simeq (14.5R_\odot, 3 \times 10^5 \K)$ and $(R,T)_ \simeq (30R_\odot, 6.4 \times 10^4 \K)$, respectively. 
We note that they disfavour a stellar source for this emission. }}}}}

We examine the following scenario in the frame of the CEJSN impostor scenario for FBOTs (\citealt{Soker2022FBOT}; section \ref{sec:intro}) to form such a source.
{{{{  While the CEJSN impostor scenario refers to the powering of the event in the first several months of the main outburst, in this study we refer to a later phase. In  the CEJSN impostor scenario the powering of the main outburst event is by a NS that accretes mass from a post-CEE circumbinary disk after the RSG envelope contracts to below the core-NS separation. This circumbinary disk survives for a timescale of $\tau_{\rm CBD} \approx 1-100~{\rm day}$ (eq. 4 in \citealt{Soker2022FBOT}). In this study we consider a later phase that occurs after the circumbinary disk does not exit anymore, namely, at about a year and later. }}}}

The basic additional assumption is that there is a low-mass {{{{ re-expanded }}}} common envelope remnant {{{{ that rebuilds itself in a time scale of several months. }}}} Either the CEE did not remove the entire RSG envelope {{{{ such that after several months of further evolution of the RSG core the RSG envelope re-expands and engulfs the NS for the second time, }}}} or after several months fallback material forms a low-density common envelope that re-engulfs the NS. {{{{{  We discussed these possibilities in section \ref{sec:intro}. We will not study them in the present paper. This paper should motivate the search for scenarios that allow a second CEE with a low-mass rebuilt envelope.  }}}}}

The NS that {{{{ is re-engulfed by the re-expanded RSG envelope, }}}} orbits now, {{{{ for the second time, }}}} inside the low density envelope remnant of the RSG star. Therefore, the NS launches jets while accreting mass from the envelope. These jets collide with the envelope and power the common envelope remnant. 
For a spherical photosphere with a uniform temperature the radius of the RSG remnant star is $R= 42 (L/10^7L_\odot)^{1/2} (T/50000\K)^{-2} R_\odot$. It is actually a blue supergiant at this phase. 

{{{{ For the rest of the paper we consider this late phase of evolution that follows the main FBOT outburst. In this phase the re-expanded low-mass RSG envelope engulfs the NS and a second phase of CEE takes place. }}}}

We consider a model where instead of a spherical uniform photosphere the star is larger and the jets heat only small polar caps. For example, for a stellar radius of $R_\ast \simeq 120 R_\odot$ an uniform effective temperature of $T_{\rm s} \simeq 30000 \K$ gives $L_\ast = 10^7 L_\odot$. Because we are limited to spherically symmetric stellar models by the numerical code (section \ref{sec:Numerical}) we cannot simulate the hot caps. Therefore, the stellar model represents a star where the zones around the equator are cooler than $T_{\rm s}$ while two hot cups spanning an angle from the pole to about $20^\circ$ from the pole are at $T_{\rm cap}=50000 \K$. 
Larger stars with even smaller caps are also possible. 

In the CEJSN impostor scenario the powering of $L>10^7 L_\odot$ comes from a NS that accretes mass via an accretion disk and launches two opposite jets.  
We follow \cite{Gricheneretal2021} and \cite{Hilleletal2022} and take for  the power of the jets that the NS launches the expression 
\begin{equation}
\dot E_{\rm 2j} =  \zeta \frac {G M_{\rm NS} \dot M_{\rm BHL,0}}{R_{\rm NS}}  ,
\label{eq:E2jzeta}
\end{equation}
where for the analytical Bondi-Hoyle-Lyttleton accretion rate we take 
\begin{equation}
\dot M_{\rm BHL,0} = \pi \rho(a) v_{\rm NS} (a)
\left[ \frac{2 G M_{\rm NS}}{v^2_{\rm NS}(a)} \right]^2 ,
\label{eq:MBHL}
\end{equation}
and where we take for the relative velocity of the NS and the envelope, $v_{\rm NS}$, the Keplerian velocity of the NS inside the envelope at its orbital radius $a$.
For the NS mass and radius we take $M_{\rm NS}=1.4 M_\odot$ and $R_{\rm NS}=12 \km$, respectively, throughout this study.

{{{{ Note that according to the CEJSN impostor scenario during the  main FBOT outburst, which lasts for up to several months, the NS accretes mass from a post-CEE circumbinary disk. Here we study a second CEE phase that occurs after the main FBOT outburst, where the low-mass RSG envelope engulfs the NS for a second time. The NS is deep inside the re-expanded envelope, and so we consider the BHL accretion process. }}}} 

Three factors determine the parameter $\zeta$ {{{{ inside a common envelope }}}} (see \citealt{Gricheneretal2021} and \citealt{Hilleletal2022} for derivation and simulations): the ratio of the accretion rate obtained in numerical simulations to that of the analytical BHL value $\dot M_{\rm BHL}$, $\xi \simeq 0.1-0.5$ (e.g., \citealt{Livioetal1986, RickerTaam2008, Chamandyetal2018, Kashietal2022}), the effect of the jets in further reducing the the mass accretion rate (the jet feedback mechanism) $\chi_{\rm j} \simeq 0.1-0.2$ (\citealt{Gricheneretal2021, Hilleletal2022}), and the ratio of the jets' power to the total accretion power $\eta \simeq 0.1$ (e.g., \citealt{Schroderetal2020}), as neutrinos carry most of the accretion energy. These values give  
\begin{equation}
\zeta \equiv \eta \chi_{\rm j} \xi \approx   10^{-3}-10^{-2} .   
\label{eq:Zeta}
\end{equation}  
Lower values are also possible, e.g., due to lower values of $\xi<0.1$ that some simulations claim for (e.g., \citealt{MacLeodRamirezRuiz2015a}; \citealt{MacLeodRamirezRuiz2015b}).  
\cite{Hilleletal2022} simulated values of $\zeta\le10^{-4}$, partly due to numerical limitations. 

As we will show in section \ref{sec:Numerical}, a typical density at several solar radii from the core is $\simeq 10^{-7} \g \cm^{-3}$. In our scenario the NS accretes mass from the envelope only as it orbits the core at several solar radii. The core mass of the models we build later is $M_{\rm core} \simeq 6.5 M_\odot$.
Substituting the mass and radius of the NS that we use here and scaling by the other variables, equation (\ref{eq:E2jzeta}) reads  
\begin{eqnarray}
\begin{aligned}
\dot E_{\rm 2j}  &  =   2.95 \times 10^7 
\left( \frac{\zeta}{0.001} \right) 
\left( \frac{M_{\rm core} + M_{\rm NS}}{8 M_\odot} \right) ^{-3/2}
\\ & \times 
\left( \frac{a}{10 R_\odot} \right)^{3/2} 
\left( \frac{\rho(a)}{10^{-7} \g \cm^{-3}} \right) 
L_\odot .
\end{aligned}
\label{eq:E2jewtsScaled}
\end{eqnarray}
The mass accretion rate by the NS is 
\begin{eqnarray}
\begin{aligned}
\dot M_{\rm NS} &  =  1.2 \times 10^{-4}  
\left( \frac{\xi \chi_{\rm j}}{0.01} \right) 
\left( \frac{M_{\rm core} + M_{\rm NS}}{8 M_\odot} \right) ^{-3/2}
\\ & \times 
\left( \frac{a}{10 R_\odot} \right)^{3/2} 
\left( \frac{\rho(a)}{10^{-7} \g \cm^{-3}} \right) 
M_\odot \yr^{-1} .
\end{aligned}
\label{eq:MnsAccretion}
\end{eqnarray}

We conclude that in principle a NS in a CEE inside the low-mass envelope remnant of a RSG that accretes mass and launches jets might power the central source that \cite{Sunetal2022} {{{{{ and \cite{Chenetal2023a}  }}}}} observe at the location of AT2018cow. At this phase the star is a blue supergiant. One outcome is that this luminosity is much above the Eddington luminosity of the system, leading to a rapid evolution. We discuss this point in section \ref{sec:Summary}. 

We end by noting that the spiralling-in process {{{{ of the NS-core binary system inside the re-expanded low-mass envelope }}}} is limited to supply the energy for a time period of $\tau_{\rm SI} < (GM_{\rm core}M_{\rm NS}/2R_{\rm core})/10^7L_\odot$. In our model the radius of the core is $R_{\rm core} \simeq 0.6 R_\odot$ (after envelope removal it expands somewhat) and we find $\tau_{\rm SI} \la 3 \yr$. It is possible that a later evolutionary phase when the core is smaller can in principle support such a powering, but we consider it less likely because the density that is required to force the NS to spiral in implies an accretion energy as well. {{{{ In any case, the re-expanded envelope is of low mass and the spiralling-in timescale is much longer than several years. }}}} This low density envelope, which leads to a very slow spiralling-in, is still sufficient to allow powerful accretion. 

\section{Numerical procedure} 
\label{sec:Numerical}

We evolve stellar models with a zero age main sequence (ZAMS) mass of $M_{\rm ZAMS}=20 M_\odot$ and with two different values of the initial metalicity $Z=0.0016$ and $Z=0.02$, with version r22.05.1 of the stellar evolution code Modules for Experiments in Stellar Astrophysics (\textsc{mesa}; \citealt{Paxtonetal2011, Paxtonetal2013, Paxtonetal2015, Paxtonetal2018, Paxtonetal2019}).\footnote{ 
The default capabilities  of \textsc{mesa} relay on the MESA EOS that is a blend of the OPAL \citep{RogersNayfonov2002}, SCVH \citep{Saumonetal1995}, FreeEOS \citep{Irwin2012}, HELM \citep{TimmesSwesty2000}, PC \citep{PotekhinChabrier2010}, and Skye \citep{Jermynetal2021} EOSes. Radiative opacities are primarily from OPAL \citep{IglesiasRogers1996}, with low-temperature data from \citet{Fergusonetal2005} and the high-temperature, Compton-scattering dominated regime by \citet{Poutanen2017}. }

At an age of $t_{\rm S}=9.68 \times 10^{6} \yr$ of the low-metalicity ($Z=0.0016$) model when the RSG core mass is $M_{\rm core,S}=6.5 M_\odot$, the RSG radius is $R_{\rm S} = 225 R_\odot$, and its luminosity is $L_{\rm S}=2.6 \times 10^{5} L_\odot$ we remove most of its hydrogen-rich envelope and leave a small remnant envelope mass $M_{\rm PS}$. This numerical rapid mass removal mimics the initial phase of the CEJSN impostor event when the NS spirals from the RSG surface to deep inside its envelope, a process that strips most of the hydrogen-rich envelope. 
We conduct the same procedure for the high metalicity ($Z=0.02$) stellar model at the age of $t_{\rm S}=8.85\times10^{6}$ when the RSG core mass is $M_{\rm core,S}=6.4 M_\odot$, its radius is $R_{\rm S} = 229 R_\odot$, and its luminosity is $L_{\rm S}=1.8 \times 10^{5} L_\odot$. 

After numerically (artificially) removing the mass we mimic the energy that the jets that the NS launches deposit to the envelope at a power of $\dot E_{2 j}=0.99 \times 10^7 L_\odot$. Together with the nuclear burning in the core the total luminosity of the star at thermal equilibrium is $L_\ast \simeq 10^7 L_\odot$. 
We deposit the energy to the envelope from an inner radius $R_{\rm Din}$ to an outer radius  $R_{\rm Dout}$, and with a constant power per unit mass. This energy deposition prescription is not unique, but we take it to reasonably describe the energy that the jets deposit in the envelope as they propagate out starting from several solar radii from the core of the giant.  

The envelope rapidly expands to a stellar radius of $R_\ast \ga 1000 R_\odot$. Due to the large radius and high luminosity the wind mass loss rate (physical mass loss rate, not an artificial one) is high and the low-mass envelope losses mass and contracts on a timescale of hundreds of years (the mass loss rate in \textsc{mesa} is based on \citealt{Glebbeeketal2009} ,\citealt{NugisLamers2000} and \citealt{Vinketal2001}). 
We present the post-stripping envelope mass $M_{\rm env}$ as function of the stellar radius $R_\ast$ for five cases in Fig. \ref{fig:MassRadius}. In Fig. \ref{fig:MassRadiusVsTime} we present the evolution of $M_{\rm env}$ and of $R_\ast$ with time $t_{\rm PS}$. The post-stripping time $t_{\rm PS}$ is measured from the end of rapid numerical mass removal.
The mass evolution of the high metalicity models is much faster than the low-metalicity models. This is expected as the mass loss rate of the high metalicity models is much higher according to the mass loss prescription. 
\begin{figure}[t]
	\centering
\includegraphics[trim=0.0cm 7.5cm 0.0cm 7.0cm ,clip, scale=0.43]{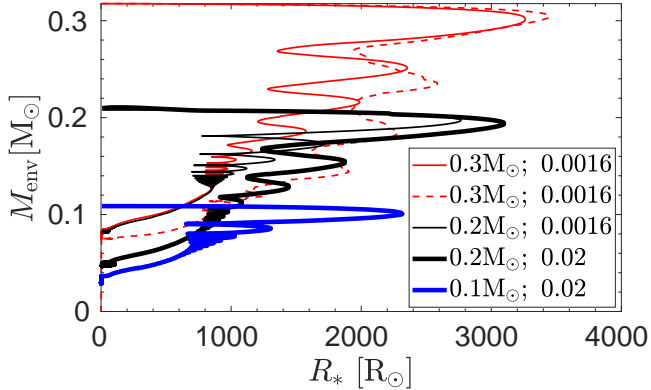}
 \\ 
\caption{Post-stripping envelope mass as function of stellar radius for five cases. The inset shows the post-stripping mass $M_{\rm env}$ and the metalicity $Z$ in the form $(M_{\rm PS}, Z)$. 
The thin and thick lines are for cases with $Z=0.0016$ and $Z=0.02$, respectively. Solid lines represent cases where we inject the energy to the envelope from $R_{\rm Din} = 1R_\odot$ to the stellar surface $R_{\rm Dout}=R_\ast$. The red-dashed line represent the case where we inject the energy between $R_{\rm Din}=1 R_\odot$ and $R_{\rm Dout} = 0.5R_\ast$. The post-stripping time $t_{\rm PS}=0$ corresponds to the end of the numerical rapid mass removal. The early phase from numerical mass removal and until the radius shrinks to $R_\ast \simeq 300 R_\odot$ is not physical and results from the numerical procedure to build the stellar model.  
}
\label{fig:MassRadius}
\end{figure}
\begin{figure}[t]
	\centering
\includegraphics[trim=2.2cm 6.0cm 0.0cm 7.0cm ,clip, scale=0.46]{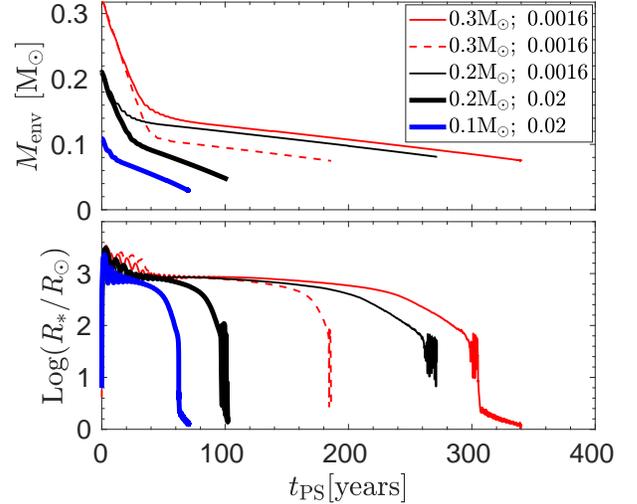} \\
\caption{The evolution with time of post-stripping envelope mass (upper panel) and the stellar radius (lower panel) after the numerical rapid mass removal. Meaning of lines is as in Fig. \ref{fig:MassRadius}. We are only interested in the models after the stellar radius is $R_\ast < 300 R_\odot$. 
}
\label{fig:MassRadiusVsTime}
\end{figure}

Our goal with the above numerical procedure is to build a stellar model with a luminosity of $L_\ast \simeq 10^7 L_\odot$ and a radius of $R_\ast \simeq 40 - 300 R_\odot$, as we described in section \ref{sec:TheScenario}. In reality there is no mass removal and then energy deposition, but rather the jets that the NS launches deposit energy during the entire CEE. 
Therefore, in what follows we deal only with the post-stripping evolutionary phase after the stellar radius becomes $R_\ast < 300 R_\odot$. 

\section{Stellar properties} 
\label{sec:Properties}

As we discuss in section \ref{sec:TheScenario} we want to present the possibility of a stellar model with a radius of $R_\ast \simeq 40 - 300 R_\odot$ and a luminosity of $L_\ast \simeq 10^7 L_\odot$. We here present the evolution of five such stellar models that we built as we describe in section \ref{sec:Numerical}. 
To facilitate comparison of the different models we set the time in what follows to be $t=0$ when the stellar radius is $R_\ast(t=0)=300 R_\odot$. 

In Fig. \ref{fig:RadiusTimeLinear} we present the evolution of the stellar radii {{{{ and luminosities }}}} with time. 
We see that the rate of radius decline depends on the metalicity and on the energy deposition prescription.
{{{{ The luminosities are negligibly smaller than $10^7 L_\odot$, showing that during the relevant phase of evolution the envelope does not add nor remove much energy as its mass and radius decrease. }}}}
\begin{figure}[t]
	\centering
\includegraphics[trim=0.0cm 5.0cm 0.0cm 2.2cm ,clip, scale=0.46]{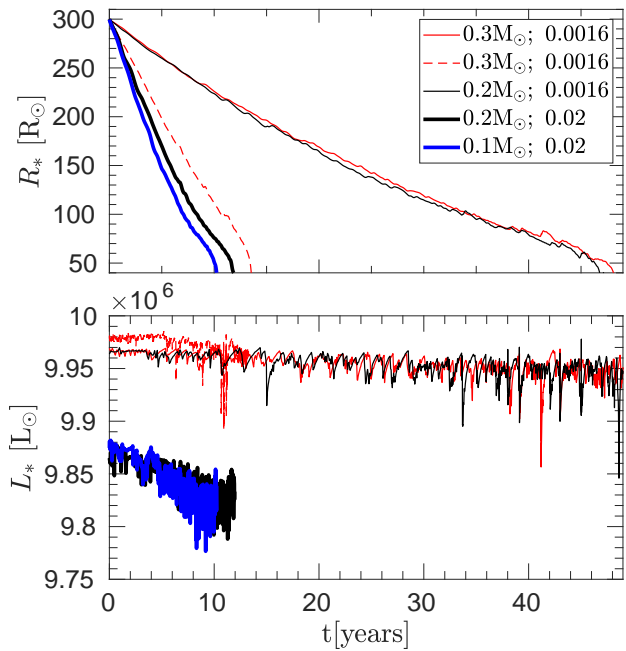}
 \\ 
\caption{Comparing the evolution with time of the stellar radii {{{{ (upper panel) and luminosities (lower panel) }}}} of the 5 models that we study here, focusing on the stellar shrinking from $R_\ast=300R_\odot$ to $R_\ast=40R_\odot$. Meaning of lines is as in Fig. \ref{fig:MassRadius}.
We arbitrarily set the time here to be $t=0$ when the stellar radii are $R_\ast(t=0)=300 R_\odot$. }
\label{fig:RadiusTimeLinear}
\end{figure}

However, even the high metalicity models (thick lines) and the model where we deposit energy in the inner half radius of the envelope (dashed line) might hold an inflated luminous envelope for several years, or even longer, if some ejected envelope gas falls back to the star. This would imply a practically lower mass loss rate. 

In Fig. \ref{fig:Density} we present the density profiles of the five models when the stellar radius of each model is $R_\ast \simeq 150 R_\odot$. We use the density $\rho(r)$ at a radius $r$ to calculate the power of the two jets that a NS at radius $r$ launches according to equation \ref{eq:E2jewtsScaled}. In Fig. \ref{fig:JetPower} we present the power of the two jets as function of the NS location and for $\zeta = 0.001$. We note that the value of $\zeta \simeq 10^{-3} - 10^{-2}$ (equation \ref{eq:Zeta}) is highly uncertain, but is expected to be in this range as we discuss section \ref{sec:TheScenario}.  
\begin{figure}[t]
	\centering
\includegraphics[trim=0.0cm 7.5cm 0.0cm 7.0cm ,clip, scale=0.43]{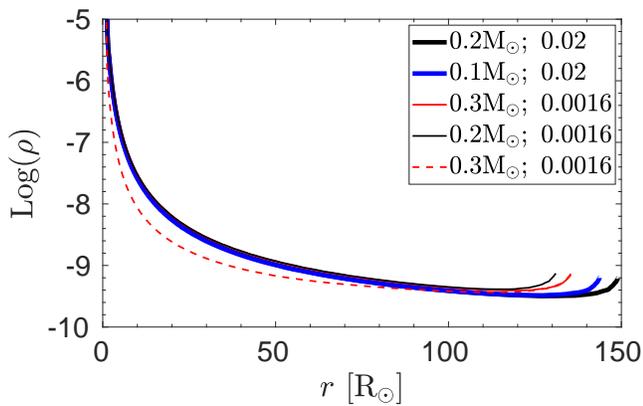}
 \\ 
\caption{The density profiles of the five models when their radii are about similar at $R_\ast \simeq 150 R_{\odot}$. Meaning of lines is as in Fig. \ref{fig:MassRadius}. 
}
\label{fig:Density}
\end{figure}
\begin{figure}[t]
	\centering
\includegraphics[trim=0.0cm 7.5cm 0.0cm 7.0cm ,clip, scale=0.43]{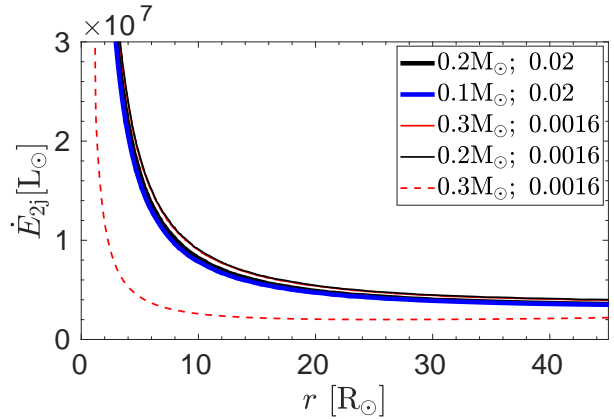}
 \\ 
\caption{The power of the two jets that the NS launches according to equation (\ref{eq:E2jewtsScaled}) with $\zeta=0.001$ as function of the NS location in the envelope for the density profiles that we present in Fig. \ref{fig:Density}. }
\label{fig:JetPower}
\end{figure}

The graphs in Fig. \ref{fig:JetPower} show that with $\zeta$ in the expected range the power of the jets can be as we take in the simulations $\dot E_{\rm 2j} \simeq 10^7 L_\odot$. In that respect our simulations are self-consistent according to the scenario we propose here.

\section{Discussion and Summary} 
\label{sec:Summary}

We conducted stellar evolutionary simulations as we described in section \ref{sec:Numerical} to build blue supergiant stellar models with a luminosity of $\simeq 10^7 L_\odot$ and radii of $R_\ast \simeq 40-300 R_\odot$. The low-mass envelope and very high luminosity mimic a possible final phase of the CEJSN-impostor scenario that we propose here (section \ref{sec:TheScenario}). This last phase of the CEJSN-impostor event (that does not happen in all cases) lasts for few years to several tens of years. During this time the NS accretes mass from the low-mass envelope and launches jets that power the system according to equation (\ref{eq:E2jewtsScaled}). 

The main findings of our simulations are as follows (section \ref{sec:Properties}). (1) We could build a stellar model with a low envelope mass of $M_{\rm env} \simeq 0.05-0.1 M_\odot$, a radius of $R_\ast \simeq 40 -300 R_\odot$, and luminosity of $L_\ast \simeq 10^7 L_\odot$ (Fig. \ref{fig:Density}).  (2) The envelope can last for few years to tens of years (Fig.  \ref{fig:RadiusTimeLinear}). This time period can be longer even if the blue supergiant accretes fallback material, implying a practically lower mass loss rate. In an extreme case the main CEE might remove the entire envelope, and a later fallback process rebuilds the low-mass envelope (we did not simulate this case here). In any case, the luminosity is highly super-Eddington and therefore this stellar model cannot last for a long time.  (3) The jets' power that we assume in building the envelope in the models, $\dot E_{\rm 2j}=0.99 \times 10^7 L_\odot$, is consistent with the expected jets' power in the CEJSN-impostor scenario as we showed in Fig. \ref{fig:JetPower}. For a given envelope structure, the jets' power depends on the location of the NS in the envelope $r$, and on the efficiency parameter $\zeta$ (equation \ref{eq:Zeta}). 

We mentioned in section \ref{sec:TheScenario} that our first motivation to study this final CEJSN-impostor phase came from the finding of \cite{Sunetal2022} of a luminous source $\simeq 2-3$ years after the outburst of the FBOT AT2018cow {{{{{ ( also \citealt{Chenetal2023a}).  }}}}}  \cite{Sunetal2022} argue that it is consistent with a hot, $T> 5 \times 10^4 \K$, and luminous, $L > 10^7 L_\odot$, black-body radiation source. However, it is not clear that this scenario is compatible with the properties of AT2018cow. The reason is that AT2018cow has X-ray variability that suggests that we do see the central engine. An optically thick giant envelope obscures the central engine. 

Nonetheless, we showed that in some cases the final phase of CEJSN-impostor events might be a blue supergiant with low-mass envelope and high-luminosity that is powered by a mass-accreting NS.

\section*{Acknowledgements}

We thank Aldana Grichener and Dmitry Shishkin for very helpful discussions and comments {{{{ and an anonymous referee for improving the manuscript. }}}} 
This research was supported by a grant from the Israel Science Foundation (769/20).

\section*{Data availability}

The data underlying this article will be shared on reasonable request to the corresponding author.  


\label{lastpage}
\end{document}